\documentstyle[aps,manuscript,amssymb]{revtex} 
\tightenlines

\begin{document} 
\draft
\title{Velocity-Dependent Friction and Diffusion for
Grains in Neutral Gases, Dusty Plasmas and Active Systems}
\author{S.A. Trigger~$^{1}$,\,\, G.J.F.van Heijst~$^{2}$,\,\, P.P.J.M.
Schram~$^{2}$}
\address{{1} Joint\, Institute\, for\, High\,
Temperatures, Russian\, Academy\, of\,
Sciences, 13/19, Izhorskaia Str., Moscow\, 127412, Russia. email:\,satrig@freenet.de}
\address{{2} Eindhoven\, University\, of\, Technology, P.O.\, Box\, 513, MB\,
5600\, Eindhoven, The\, Netherlands}
\maketitle
\begin{abstract}
A self-consistent and universal description of friction and
diffusion for Brownian particles (grains) in different systems,
as a gas with Boltzmann collisions, dusty plasma with ion
absorption by grains, and for active particles (e.g., cells in
biological systems) is suggested on the basis of the appropriate
Fokker-Planck equation. Restrictions for application of the
Fokker-Planck equation to the problem of velocity-dependent
friction and diffusion coefficients are found. General description
for this coefficient is formulated on the basis of master
equation. Relation of the diffusion coefficient in the coordinate
and velocity spaces is found for active (capable to transfer
momentum to the ambient media) and passive particles in the
framework of the Fokker-Planck equation.

The problem of anomalous space diffusion is formulated on the
basis of the appropriate probability transition (PT) function. The
method of partial differentiation is avoided to construct the
correct probability distributions for arbitrary distances, what is
important for applications to different stochastic problems.
Generale equation for time-dependent PT function is formulated and
discussed.

Generalized friction in the velocity space is determined and
applied to describe the friction force itself as well as the drag
force in the case of a non-zero driven ion velocity in plasmas.
The negative friction due to ion scattering on grains exists and
can be realized for the appropriate experimental conditions.
\end{abstract}
\section{Introduction}
Interest in Brownian dynamics is conditioned by a large variety of
applications: granular systems, including dusty plasmas, various
objects in biological systems, physical-chemical systems, et
cetera. For open systems and for systems with non-elastic
processes the relations between the friction and diffusion
coefficients and even the correct specific forms of the
Fokker-Planck equation are not still established. Recently the
friction and diffusion coefficients as functions of the grain
velocity $V$ were derived for dusty plasmas \cite{SSTZ2000}. Due
to ion absorption by the grains, the friction coefficient can
become negative \cite{TZ2002} for the simplest model of dusty
plasmas with ion absorption without atom regeneration on the
grain's surface. This implies that the problem of negative
friction has to be considered for more complicated models and
realistic situations in dusty plasmas, in particular taking into
account the grain's mass conservation for, which is essential for
large times of the process, in spite of big difference between ion
and grain masses. In the present paper we consider the specific
forms of the probability transition (PT) \cite {T2002} for
Boltzmann-type and and non-Boltzmann-type of collisions to
calculate and compare the velocity dependent friction and
diffusion coefficients in different systems, including gases,
dusty plasmas and active particles. Our main goal in this paper is
determination of the velocity-dependent friction and diffusion
coefficients for arbitrary grain velocity and consideration of
diffusion in the coordinate and velocity spaces for different
systems and conditions.

In Section 2 the Fokker-Planck equation and self-consistent
description of velocity-dependent friction and diffusion are
discussed in details on basis of the PT function (PT). In Section
3 we consider in detail the probability transition function for
Boltzmann-type collisions, in particular, for the Boltzmann's
spheres. Friction and diffusion coefficients are calculated on the
basis of the Fokker-Planck equation. The restriction for
application of the Fokker-Planck equation to the fast motion of
grains for the Boltzmann's type of collisions and necessity to use
the appropriate master equation with prescribed PT function is
shown. The general formulae for the velocity-dependent friction
and diffusion coefficients in the velocity space are established
on the basis of master equation. In Section 4 the absorption
collisions for ions by grains is done for dusty plasmas with a
high grain charge. In the framework of the simplest model with a
fixed mass of grains the phenomena of negative friction is found.
The friction coefficient changes sign for some value of the grain
velocity. Manifestation of negative friction due to ion absorption
in dusty plasmas is the consequence of the model. Opportunity to
realize this process in the experiment can be considered on more
elaborated model with grain mass conservation. In Section 5 we
consider diffusion in the coordinate space on the basis of the
Fokker-Planck kinetic equation. The general relation between the
diffusion coefficient in the coordinate space and
velocity-dependent diffusion and friction coefficients in the
velocity space are found. Relation of these results with the known
limiting cases is established. The results are applied to the
self-motion of grains (e.g. cells) in biological systems. In
Section 6 the different types of anomalous diffusion in the
coordinate space are considered on the basis of the appropriate
master equation. As particular case we describes the Levy-flights,
which are very important for many applications. We also formulated
the generalized diffusion equation for the time-dependent PT
function in the coordinate space, investigated it for some
particular cases and suggest to apply for subdiffusion processes.
In Section 7 we shortly describe the generalized friction, which
appears when some other vectors, except grain velocity, are
essential to determination of PT function in the velocity space.
The application of such consideration is done in Section 8, where
the ion flow with some driven velocity (which is the particular
case of such additional vector in PT function) scatters on grains.
Due to this process the total friction can be negative for small
grain velocity and provides the mechanism for acceleration and
heating grains in dusty plasmas. The conclusions are presented in
the Section 9.

\section{Fokker-Planck equation and self consistent
velocity-dependent friction and diffusion}
The friction and diffusion coefficients $\beta(V)$,
$D_\parallel(V)$ and $D_\perp (V)$ on the basis of the
Fokker-Planck equation are determined via PT function $w({\bf P,
q})$ \cite {T2002} as
\begin{eqnarray}
\beta(V)=\frac{1}{P^2}\int d\,^s{\bf q}\, ({\bf Pq})\, w({\bf
P,q}),\label{be0}
\end{eqnarray}
\begin{eqnarray}
D_\parallel(V)= (1/2M^2 P^2) \int d\,^s{\bf q}\,{\bf (Pq)}^2
w({\bf P, q}). \label{be1}
\end{eqnarray}
and
\begin{eqnarray}
D_\perp(V) = \frac{1}{2(s-1)M^2 P^2} \int d\,^s {\bf q}\,\left
[P^2 q^2 - {\bf (Pq)}^2\right] w({\bf P, q}). \label{bea2}
\end{eqnarray}
Here ${\bf (Pq)}$ is the scalar product in velocity space of the
dimension $s$ and ${\bf P} \equiv M{\bf V}$, where $M$ and $V$ are
respectively the mass and velocity of the grain.

We determine the PT function $w({\bf P, q}) \equiv w({\bf P,
P'-P})$ (${\bf q \equiv P'-P}$ represents the momentum transfer for
the considered process, e.g. collision) in the same way
(convenient for consideration of the Fokker-Planck equation) as
was done in \cite {LP}:
\begin{eqnarray}
\frac{df_g({\bf P},t)}{dt} =  I_g({\bf P},t) = \int d{\bf q}
\left\{w({\bf P+ q, q}) f_g({\bf P+q, t}) \right.\nonumber\\
\left.- w({\bf P, q}) f_g({\bf P},t) \right\}, \label{dust}
\end{eqnarray}
The corresponding to Eq.~(\ref{dust}) Fokker-Planck equation can
be written as:
\begin{eqnarray}
\frac{d f_g({\bf P},t)}{d t} = {\partial \over {\partial P_i}}
\left[ P_i \beta({\bf P}) f_g({\bf P}) + M^2 {\partial \over
{\partial P_j}} \left( D_{ij}({\bf P}) f_g({\bf P}) \right)\right]
\label{FP}
\end{eqnarray}
The PT function $w({\bf P, q})$ is directly connected with the
probability transition $w_0({\bf P, P'})$ in the
Chapman-Kolmogorov equation (master equation, \cite{RL}) $w_0({\bf
P',P})= w({\bf P, P-P')}\equiv w({\bf P, q})$:
\begin{eqnarray}
\frac{df_g({\bf P},t)}{dt} = \int d{\bf P'} \left\{w_0({\bf P,
P'}) f_g({\bf P', t}) - w_0({\bf P', P}) f_g({\bf P},t) \right\}.
\label{CK}
\end{eqnarray}
It follows from Eqs.~(\ref{be0}), (\ref{be1}) that the friction
and diffusion coefficients always have to be calculated
\emph{self-consistently}. For the Boltzmann-type collisions
between the light particles (conventionally atoms) and grains with
masses $m$ and $M$ ($m\,\ll\,M$), respectively, the PT function
$w({\bf P, q})$ and the coefficients $\beta(V)$ and $D(V)$ can be
found for arbitrary cross-section (as will be done in Section 2).
In Section 2 we also investigate the applicability of the Einstein
relation to the Boltzmann-type collisions on the basis of the
Fokker-Planck equation with velocity-dependent coefficients.

For some other type of collisions the problem of a single-valued
form of $w({\bf P, q})$ exists, because the exact kinetic equation
for such collisions is not known in many cases. The Einstein
relation between friction and diffusion of grains can be easily
violated for such type of non-Boltzmann collisions, even for very
low grain velocity.

The solution of the stationary Fokker-Planck equation for the
isotropic distribution function of the particles can be written
\cite {T2002} as:
\begin{equation}
f_g(V) =  \frac{C} {D_\parallel(V)} \exp\left[-\int\limits_0^V
d\upsilon \upsilon \frac
{\beta^\ast(\upsilon)}{D_\parallel(\upsilon)}\right], \label {fgr}
\end{equation}
where $C$ is a constant, providing normalization and $D_\parallel$
is the longitudinal part of the diffusion tensor:
\begin{equation}
D_{ij}({\bf P}) = \frac{P_i P_j}{P^2}D_\parallel(P)\ + \,
(\delta_{ij} - \frac{P_i P_j}{P^2})\, D_\perp(P).\label{scul}
\end{equation}
The function $\beta^\ast(V)$ is related with $\beta(V)$ by the
equality:
\begin{equation}
\beta^\ast(V)=\beta(V)+ \frac{s-1}{V^2}\left (D_\parallel(V) -
D_\perp(V)\right) .\label{scul1}
\end{equation}
From Eq.~(\ref{fgr}) follows:
\begin{eqnarray}
\beta^\ast (V)= -\frac{1}{V} \frac{\partial D_\parallel}{\partial
V}-\frac{D_\parallel}{V f_g} \frac{\partial f_g}{\partial V}.
\label{be}
\end{eqnarray}
Here the function $f_g$ is the stationary grain distribution function.

\section{Boltzmann-type collisions: limitations of the
Fokker-Planck equation. Master equation approach}
For the Boltzmann-type collisions the stationary distribution
$f_g(V)$ is a Maxwellian equilibrium $f_{g0}(V)$ and
Eq.~(\ref{be}) has the form
\begin{eqnarray}
\beta (V)= -\frac{1}{V} \frac{\partial D_\parallel(V)}{\partial
V}+\frac{M D_\parallel(V)} {T},\label{bee}
\end{eqnarray}
If $D$ is a function of $V$  the Einstein relation $\beta=MD/T$ is
violated if the term with derivative in Eq.~(\ref{bee}) is not
small with respect to the parameter $m/M$. Only in that special
case deviation from the Einstein relation in Eq.~(\ref{bee}) can
be neglected, according to the general conditions of applicability
for the Fokker-Planck equation. For Boltzmann collisions it is
possible to show that the PT function $w_B({\bf P, q})$ (with
${\bf P}\equiv{\bf P_a}=M{\bf V}$) can be represented as:
\begin{eqnarray}
w_B({\bf P, q})= \frac{1}{\mu^2} \int d{\bf u}\, \delta \left(
{\bf u q} - \frac{q^2}{2\mu} \right)\cdot \frac{d \sigma}{do}
\left [\arccos\, (1-\frac{q^2}{2\mu^2 u^2}), u \right] f_b ({\bf
u+V-v_0}) .\label{cf1}
\end{eqnarray}
Here $\mu$ is the reduced mass of the particles of species $a$ and
$b$ (with masses $M$ and $m$, respectively), $d\sigma/do$ is the
differential cross-section for Boltzmann-type scattering and $f_b$
is the velocity distribution function for the light particles,
which we assume to be Maxwellian with, for generality, some
shifted velocity. It means small particles can have some driven
velocity ${\bf v_0}$.

For the case of hard spheres, when $d\sigma/do = a^2/4$ and the
distribution function of the small particles $f_b=f_{b0}$ is
Maxwellian , Eq.~(\ref{cf1}) can be written as:
\begin{eqnarray}
w_B^0({\bf P, q})= \frac{a^2 n_b}{4q \mu^2} \left(\frac{m_b}{2\pi
T_b}\right)^{1/2} \exp\left [-\frac{m_b}{2T}\left(\frac{\bf V
q}{q}-\frac{q}{2\mu}\right)^2 \right],\label{cf}
\end{eqnarray}
where $a$ is the reduced radius of the colliding spheres. Let us
calculate the friction and diffusion coefficients on the basis of
general equations (\ref{be0}),(\ref{be1}) and the PT function
(\ref{cf}) for hard spheres. It is  easy to show that the friction
coefficient can be represented via the function of parabolic
cylinder $D_{-3}$ in the form:
\begin{eqnarray}
\beta(V)= \beta(0)\,I\left(\frac{V}{v_T}\right),\;\; I \equiv
\frac{3}{2 \omega}\int_{-1}^{1}dx x \exp \left[- \frac {\omega^2
x^2}{2}+ \frac {\omega^2 x}{4}\right] D_{-3}(-\omega
x),\label{be2}
\end{eqnarray}
where $\omega\equiv V/v_T$ and $v_T=\sqrt{T/m}$ is the thermal
velocity of the small particles. The value of the friction
coefficient for the grain in the hard spheres media for $V=0$ is
equal:
\begin{eqnarray}
\beta(0)=8\sqrt{2\pi}\,\mu n a^2 v_T /3M. \label{beb2}
\end{eqnarray}
Likewise for the friction coefficient we derive
\begin{eqnarray}
\emph{D(V)}= \emph{D}(0)J \left(\frac{V}{v_T}\right), \;\,J=
\frac{9}{2}\int_{-1}^{1} dx x^2 \exp \left[- \frac {\omega^2
x^2}{2}+ \frac {\omega^2 x}{4}\right] D_{-4}(-\omega x).
\label{be3}
\end{eqnarray}
For $V=0$ the coefficients $\emph{D}(0)$ and $\beta(0)$ are
related ($\mu \simeq m$) by Einstein's law:
\begin{eqnarray}
\emph{D}(0)=\mu T \beta(0)/ m M. \label{be4}
\end{eqnarray}
For the opposite case of very high velocity $V \gg v_T$ the formal
expressions for the coefficients $\beta$ and $D$, based on
self-consistent relations (\ref{be0}),\,(\ref{be1}) and the PT
function for Boltzmann-type collisions (\ref{cf1}) are:
\begin{eqnarray}
\beta (V \gg v_T)=\frac{\pi\mu}{M} a^2 n V, \label{be5}
\end{eqnarray}
\begin{eqnarray}
\emph{D}(V \gg v_T)= \beta(V \gg v_T)\frac {2\mu V^2}{3M}=
\frac{2\pi\mu^2}{3M^2} a^2 n V^3 \label{be6}
\end{eqnarray}
The similar expressions have been found in \cite{T2002} (although
here we corrected the numerical multipliers order of unit) and
evidently violate the Einstein relation. As it follows from
Eq.~(\ref{cf}) the conditions of applicability of the
Fokker-Planck equation cannot be restricted only by the usual
inequality (e.g. \cite {LP}) ${\bf q \ll P}$. To find the full
limitations for transition from the Boltzmann equation to the
Fokker-Planck equation, let us consider the conditions for
expansion of the function $w({\bf P+q, q})$ by ${\bf q}$ in
Eq.~(\ref{dust}) for the simple particular case described by
Eq.~(\ref{cf}):
\begin{eqnarray}
w_B^0({\bf P+q, q})= w_B^0({\bf P,q})\exp\left
[-\frac{m_b}{2T}\left(\frac{\bf V
q}{q}-\frac{q}{2\mu}\right)\frac{2q}{M}-\frac{m_b}{2T}\frac{q^2}{M^2}
\right].\label{cf2}
\end{eqnarray}
To obtain the Fokker-Planck equation from the master equation
~(\ref{dust}) we have to expand the exponent in Eq.~(\ref{cf2}).
Therefore besides the general condition ${\bf q \ll P}$ the
inequalities $\mid {\bf V}\mid\ll v_T^2 M/q$ and $q^2/TM \ll 1$
have to be fulfilled simultaneously. An even more stronger
limitation for applicability of Fokker-Planck equation for
Boltzmann-type collisions $\mid {\bf V}\mid \ll T/ q$ follows from
the fact that the stationary state of the distribution function
for grains (without external fields) is Maxwellian. This means
that for any finite $q$ there is some limitation on the values
$\mid {\bf V}\mid$, which can be described by the Fokker-Planck
equation. In particular, according these inequalities the limit
$T\rightarrow0$ cannot be realized for the Boltzmann-type
collisions on the basis of the Fokker-Planck approximation
simultaneously with finite $\mid {\bf V}\mid$-values. In other
words, the expressions for the friction and diffusion coefficients
(\ref{be5}),\,(\ref{be6}) for a Boltzmann gas for the velocities,
which are not satisfied the inequalities, mentioned above, are in
contradiction with Eq.~(\ref{bee}).

Because the Fokker-Planck equation itself is not valid in
Boltzmann gas for high velocities, Eqs.~(\ref{FP}), (\ref{bee}) is
not applicable and our aim is to find the friction and diffusion
coefficients for arbitrary velocities on the basis of the
Chapman-Kolmogorov equation (\ref{CK}). For this equation in the
absence of the external fields the stationary distribution for
grains is, as easy to see, the Maxwellian if the transfer
probability function is calculated with the Maxwellian
distribution of the light particles (\ref{cf}). Equation
(\ref{CK}) permits to determine in a physically justified way the
friction and diffusion coefficients for arbitrary particle
velocities in the same form (\ref{be0}),\,(\ref{be1}) as it was
found from the Fokker-Planck equation, in spite the equation
itself is not valid anymore.

In order to show this let us integrate equation (\ref{bee}) with
the multiplayer $P_\alpha$ and with the product $P_\alpha
P_\beta$. Then we arrive to the equations for the momentum:
\begin{eqnarray}
\frac{d \langle {\bf P} \rangle}{dt} = - \int d {\bf P} f_g ({\bf
P}) \int d {\bf P'} w_0({\bf P', P}) {\bf (P-P')} = \nonumber \\
- \int d {\bf P} f_g (P) \int d {\bf q} w({\bf P, q}) {\bf q}
\equiv - \int d {\bf P} f_g (P){\bf P} \beta ({\bf P}). \label{M1}
\end{eqnarray}
Here $\langle{\bf P}\rangle$ is the momentum, averaged by the
distribution function $f_g$ and $\beta ({\bf P})$ is exactly the
same as in Eq.~(\ref {be0}). Equation for the averaged product
$\langle{\bf P}_\alpha {\bf P}_\beta \rangle$ is:
\begin{eqnarray}
\frac{d \langle {\bf P_\alpha P_\beta}\rangle} {dt} = - \int d
{\bf P} f_g ({\bf P}) \int d {\bf P'} \left({\bf P}_\alpha {\bf
P}_\beta - {\bf P'}_\alpha {\bf P'}_\beta \right) w_0({\bf P',
P}) \equiv \nonumber\\
\int d {\bf P} f_g ({\bf P})\left[2 \frac{P_\alpha
P_\beta}{P^2}\left (M^2 D_\parallel(P) -P^2 \beta (P)\right)  + 2
M^2\, \left(\delta_{\alpha\beta} - \frac{P_\alpha
P_\beta}{P^2}\right)\, D_\perp(P)\right]\equiv \nonumber\\
2 M^2 \int d {\bf P} f_g ({\bf P})\left[ D_{\alpha \beta}({P})-
\frac{P_\alpha P_\beta}{M^2}\beta (P)\right]  , \label{M2}
\end{eqnarray}
The function $D_\parallel(P)$ $D_\perp(P)$ are the same as in
Eqs.~(\ref{be1}), (\ref{bea2}), which were found from the
Fokker-Planck equation, although this equation is strictly not
applicable and we derived Eqs.~ (\ref{M1}), (\ref{M2}) on the
basis of master equation (\ref{CK}).

If we recalculate the time derivatives for the averaged momenta on
the basis of Fokker-Planck equation (\ref{FP}) we would find the
same form of the resulting equations (\ref{M1},\ref{M2}).
Naturally, the function $f_g$ in both cases is different and
satisfies to the relevant (Chapman-Enskog or Fokker-Planck)
equations. It means that determination of the friction and
diffusion coefficients by Eqs.~ (\ref{be0})-(\ref{bea2}) is
general for the system of a two components, if the distribution
function for the light component is prescribed.  In this way we
express the equation for the momentum of distribution function
with the general velocity-dependent friction and diffusion in the
velocity space. Of cause, the relation (\ref{be}) is not valid
anymore, nor are the Einstein relations for the velocity dependent
coefficients. This consideration justifies and generalizes the
statements of \cite{T2002}. Interpolation relation between the
friction and diffusion coefficients, which is valid for all
velocities for the considered case can be written in the form:
\begin{equation}
D(V)= \frac{T}{M} \left(1 + \frac{2m V^2}{3T}\right) \,\beta (V).
\label{Dq}
\end{equation}
which corresponds with the Einstein relation only for the case
that the velocities are much smaller than thermal one $V\ll v_T$.
In this sense a discussion of the relations between the
velocity-dependent friction and diffusion coefficients on the
basis of the Fokker-Planck equation \cite{KL} is relevant for
non-Boltzmann collisions; for Boltzmann-type collisions this
discussion seems artificial, if there is not some specific
additional parameter, which makes the condition ${\bf q\ll P}$
sufficient for using of the Fokker-Planck equation, as it is for
the case of collisions in a plasma with Coulomb interaction.

At the same time in parallel with the interpolation relation
(\ref{Dq}), which, naturally, is not single-valued, the
\emph{exact} integral relation between the velocity dependent
friction and diffusion exists as it follows from Eq.~(\ref{M2}):
\begin{equation}
\int d {\bf P} f_{g0} ({\bf P}) D_{\alpha \gamma}({\bf P})= \int d
{\bf P} f_{g0} {\bf P_\alpha}{\bf P_\gamma}\, \beta ({\bf P}).
\label{E1}
\end{equation}
This relation is the integral generalization of the Einstein's law
and transfers to it for the case of velocity-independent friction
and diffusion coefficients.

\section{Absorption  collisions}
For absorption collisions (typical, e.g., for dusty plasmas) the
momentum transferred to the grain is equal to the momentum of the
colliding atom (ion in the case of dusty plasmas) ${\bf p}$. The
PT function\, $w({\bf P, q})$ in this case can be found
\cite{T2002} on the basis of the kinetic equation for absorption
\cite{IG1998}
\begin{eqnarray}
w({\bf P, q})=  f_n({-{\bf q}}) \sigma_{abs} \left(\left|\frac{\bf
P}{M} +  \frac{\bf q}{m}\right| \right) \left| \frac{\bf P}{M}
+\frac{\bf q}{m} \right|. \label{w4}
\end{eqnarray}
For the Maxwellian distribution of atoms and purely geometrical
absorption $\sigma_{abs} = \pi a^2$ \,\,(the particular case of
the absorption cross-section for the model of dusty plasmas with
the average charge of grain $Q\rightarrow0$), we find for
arbitrary $V$:
\begin{eqnarray}
\emph{D(V)}= 2M^{-1}T \beta(V).\qquad \label{E1}
\end{eqnarray}
For $V=0$ this result was found in \cite{SSTZ2000}. Equation
(\ref{E1}) is different from the Einstein relation because the
absorption collision integral is different from the
Boltzmann-type.

For the real dusty plasmas, where the average charge of grain $Q$
is very high (order of $10^3\div10^4$ of the electron charge) the
absorption cross-section can be taken in so called orbital motion
limited (OML). This approach based on the assumption that there is
no potential barrier for the ions moving towards the grain and the
conservation of angular momentum and energy is enough to obtain
the cross-section $\sigma_{abs}\equiv\sigma_{OML}$ of absorption
for the charge $e_\alpha$ by grain with the radius $a$:
\begin{eqnarray}
\sigma_{OML}(Q,v_\alpha)&=& \pi a^2 \left(1-{2e_\alpha Q\over
m_\alpha v_\alpha^2a}\right) \theta \left(1-{2e_\alpha Q\over
m_\alpha v_\alpha^2 a}\right).\qquad \label{E2}
\end{eqnarray}
Here $v=|{\bf v}_\alpha-{\bf V}|$ is the relative velocity of the
colliding particles. As it follows from Eqs.~(\ref{E1}),(\ref{E2})
the Fokker-Planck equation is always applicable to the absorption
collisions under consideration if $P \gg q$ due to smooth
dependence of the PT function from the momentum $P$. Calculation
of the velocity-dependent friction and diffusion coefficients on
the basis of PT function (\ref{w4}) and
Eqs.~(\ref{be0})-(\ref{bea2}) leads to the conclusion, that
friction coefficient in the model of dusty plasmas with dominant
absorption collisions always negative in some velocity domain if
the parameter of ion-grain interaction $\Gamma\equiv Q e_i /a T_i$
\cite{TZ2002}.For the range of parameters $\eta\equiv
V^2/v_{Ti}^2\ll1$ and $1\leq\Gamma<2$ the absorption friction
coefficient for ions $\beta_i$ has a form:
\begin{eqnarray}
\beta_i(\eta,\Gamma)=2A\left[
1-\Gamma-{\eta\over5}(1-3\Gamma)\right],\label{E3}
\end{eqnarray}
and negative for $\eta\leq 5(\Gamma-1)/(3\Gamma-1)$. The
coefficient $A$ is equal:
\begin{eqnarray}
A={1\over3}\sqrt{2\pi} \left({m_i\over m_g}\right)a^2n_iv_{\rm
Ti}. \label{E4}
\end{eqnarray}
For  $\eta\gg 1$ and arbitrary values of $\Gamma$ the equation
(\ref{be0}) gives:
\begin{eqnarray}
\beta_i(\eta,\Gamma)={3\over2} A\sqrt{{\pi\over\eta}}\left[
1-{1+2\Gamma\over 2\eta}\right]\label{E5}.
\end{eqnarray}
It means that for $\Gamma\gg1$ and large $\eta$ the equation
$\beta_i(\eta_1\Gamma)=0$ has  a root $\eta_1(\Gamma)\simeq
(1+2\Gamma)/2\sim\Gamma$. Naturally, for all $\Gamma>1$ there
exists appropriate $\eta(\Gamma)$, which is the  root of the
equation $\beta_i(\eta,\Gamma)=0$.

Necessary to mention that it does not mean that the total friction
in dusty plasmas can be negative, because there are other
mechanisms of positive friction (e.e. atomic and ion scattering by
grains) which provide usually totally positive friction for
grains. Manifestation of negative friction due to absorption
requires the very special condition. In particular we have stress
that the friction coefficient $\beta_i$, which we discussed is
finite for all values of velocities. Below we consider another
mechanism of negative friction, for which always exist the
experimental conditions.

We have to stress that Eq.~(\ref{w4}) describes the absorption
collisions without mass conservation, otherwise mass of the grains
will increase and stationary solution does not exist
\cite{Tr01}-\cite{ITME02}. In real dusty plasma both cases
(stationary and non-stationary regimes) are realized, depending
from the types of plasma, discharges and parameters. For the case
of stationary solutions the mass conservation provides by surface
recombination of ions and remove of the creating atoms back to
plasma. Naturally this more complicated situation of absorption
with further atom emission by grain cannot be described by PT
function (\ref{w4}) and requires a special consideration.
Therefore the mechanism of negative friction considered above has
mostly illustrative interest as an example of microscopical model
for negative friction.

\section{Diffusion in the coordinate space:
self-consistent description for the Fokker-Planck approximation}
Let us consider the relation of diffusion in coordinate space with
diffusion in the velocity space on the basis of the Fokker-Planck
equation. It means as it was shown above, that for the case of
Boltzmann gas we have consider a low grain velocity (with the
velocity independent friction and diffusion coefficient, when the
Fokker-Planck equation is applicable). At the same time for the
non-Boltzmann type of collisions, such a type as the absorption
collisions in plasmas or active motion of biological objects (see
\cite{T2002} and the analysis below) we can consider the
coefficients, which are essentially dependable from the velocity,
because the Fokker-Planck equation in that case based on the
processes of non-Boltzmann type. The problem of relation between
the diffusion in the coordinate and velocity spaces date to
Chandrasekhar \cite {CS} has been considered in a few papers (e.g.
\cite{SG93}-\cite{SFG94}) for the Fokker-Planck equation, but
still is not solved self-consistently. To find the diffusion $D_R$
in the coordinate space we have consider a slightly inhomogeneous
in space stationary solution for the Fokker-Planck equation (see
inhomogeneous variant of Eq.~(\ref{FP}):
\begin{eqnarray}
{\bf v} \frac {\partial f_g ({\bf P, r},t)}{\partial {\bf r}} =
{\partial \over {\partial P_i}} \left[ P_i \beta(P)f_g({\bf P,
r},t)  + M^2 {\partial \over {\partial P_j}} \left( D_{ij}({\bf
P}) f_g({\bf P, r},t) \right)\right] \label{FP1}
\end{eqnarray}
For the distribution function one can use the expansion $f_g({\bf
p,r},t)=f_g^0({\bf p,r},t)+ f_g^1({\bf p,r},t)$, where $f_g^0$ is
has the same shape as the stationary homogeneous solution, but
with coordinate-dependent density $n_g ({\bf r})$ and temperature
$T_g ({\bf r})$. The equation for perturbation $f_g^1$ can be
written in the form:
\begin{eqnarray}
P_i \tilde\beta(P)f_g^1({\bf P, r},t) + M^2 \left( D_{ij}({\bf
P}){\partial \over {\partial P_j}} f_g^1({\bf P, r},t) \right)=
{\partial \over {\partial r_j}} \Phi (P, {\bf r})\equiv {\bf
\Psi}_j ( P,{\bf  r}), \label{FP2}
\end{eqnarray}
where the functions $\tilde\beta$ and $\Phi$ are equal:
\begin{eqnarray}
\tilde\beta(P) = \beta(P)+ \frac{s-1}{V^2}\left (D_\parallel(V) -
D_\perp(V)\right) + \frac{1}{V} {\partial D_\parallel(V)\over
{\partial V}}\equiv \beta^\ast(V)+\frac{1}{V} {\partial
D_\parallel(V)\over {\partial V}},   \label{B2}
\end{eqnarray}
\begin{eqnarray}
\Phi (P, {\bf r})= - \frac{1}{M}\int_P^\infty d P' P' f_g^0(P',
{\bf r}). \label{B3}
\end{eqnarray}
Because for the slight inhomogeneity one can represent the
function ${\bf \Psi}$ as the linear combination of the gradients:
\begin{eqnarray}
{\bf \Psi} (P, {\bf r})= \lambda (P)\nabla n ({\bf r})+\gamma
(P)\nabla T ({\bf r}), \label{B5}
\end{eqnarray}
it is natural to find a solution $f_g^1({\bf P,r},t)$ of Eq.~
(\ref{FP2}) as
\begin{eqnarray}
 f_g^1(P, {\bf r})= \mu_1 (P){\bf P}\cdot\nabla n ({\bf
 r}) + \mu_2 (P) {\bf P}\cdot \nabla T ({\bf r}). \label{B6}
\end{eqnarray}
Taking into account that $f_g^0 (P)$ is always proportional to the
density of grains $n({\bf r})$ we easily determine the functions
$\lambda(P)$ and after that $\mu_1(P)$:
\begin{eqnarray}
\lambda(P) = - \frac{1}{M n} \int_P^\infty d P'P'f_g^0(P'),
\label{B7}
\end{eqnarray}
\begin{eqnarray}
\mu_1(P) = \frac{1}{M^2P} \int_0^P d P'
\frac{\lambda(P')}{D_\parallel(P')}\exp \left[\int_P^{P'} d
P''\tau(P'') \right]. \label{B8}
\end{eqnarray}
The function $\tau(P)$ under the integral in the exponent of
Eq.~(\ref {B8}) is equal:
\begin{eqnarray}
\tau(P)= \frac{P \tilde \beta(P)}{M^2 D_\parallel(P)}. \label{B9}
\end{eqnarray}
For self-consistent description according to Eq.~(\ref {fgr}) we
can rewrite $\mu_1(P)$ as:
\begin{eqnarray}
\mu_1(P) = \frac{f_g^0(P)}{M^2P} \int_0^P d P'
\frac{\lambda(P')}{D_\parallel(P')f_g^0(P')}\equiv
-\frac{f_g^0(P)}{M^3nP} \int_0^P d P'
\frac{1}{D_\parallel(P')f_g^0(P')}\int_{P'}^\infty d \tilde P
\tilde P f_g^0(\tilde P). \label{Ba8}
\end{eqnarray}
The same way can be found $\mu_2(P)$.

Because the diffusion current is determined by the relation:
\begin{eqnarray}
{\bf i}_D = \int d {\bf P} {\bf V} f_g^1(P, {\bf r})\equiv -D_R
\nabla \cdot n(\bf r), \label{B10}
\end{eqnarray}
we find straightforward in the framework of the Fokker-Planck
equation the general relation:
\begin{eqnarray}
D_R = -\frac {1}{s M}\int d^s P P^2 \mu_1 (P)\equiv \frac{1}{M^4 s
n} \int d^s P P f_g^0(P)  \int_0^P d P'
\frac{1}{D_\parallel(P')f_g^0(P')}\int_{P'}^\infty d \tilde P
\tilde P f_g^0(\tilde P), \label{B11}
\end{eqnarray}
where $s=3$ for tree-dimensional case and $s=2$ for
two-dimensional case. The values $\beta(P)$, $D_\parallel$ and
$D_\perp$ are determined self-consistently by the relations
(\ref{be0}),(\ref{be1}) and (\ref{bea2}) via the probability
transition function. Introducing the characteristic momentum
$P_0$, velocity $V_0$, and constant friction $\beta_c$ and
diffusion $D_c$ coefficients by formulae $P_0=MV_0$,
$\beta(P)=\beta_c b(P)$ and $D(P)=D_c d(P)$ let us rewrite
Eq.(\ref{B11}) in undimensional form:
\begin{eqnarray}
D_R = \frac{V_0^4}{s D_c} \int d^s \eta \eta \tilde f_g^0(\eta)
\int_0^\eta d \gamma \frac{1}{d_\parallel(\gamma)\tilde
f_g^0(\gamma)}\int_{\gamma}^\infty d \omega \omega \tilde
f_g^0(\omega),\label{Ba11}
\end{eqnarray}
where the distribution function $\tilde f_g^0(\eta)$ ($\eta\equiv
P/P_0$) is normalized to unit:
\begin{eqnarray}
\int d^s \eta \tilde f_g^0(\eta)\equiv\int d^s \eta
\frac{1}{d\parallel(\eta)}\exp \left[- \Lambda\int_0^\eta d\zeta
\zeta \frac {b(\zeta)}{d_\parallel(\zeta)} \right]=1. \label{Bb11}
\end{eqnarray}
The parameter $\Lambda\equiv \beta_c V_0^2/D_c$. It means in
general the relation of the diffusion coefficients in the
coordinate and velocity space (\ref{B11}) can be written as:
\begin{eqnarray}
D_R = \frac {V_0^4}{s D_c}\,\digamma \left( \frac{\beta_c
V_0^2}{D_c}\right),
\label{Bc11}
\end{eqnarray}
where we determine the integrals in Eq. (\ref{B11}) as
$\digamma\left( \beta_c V_0^2 /D_c \right)$.

Let us consider the particular cases.\\
a) For the constant values of $\beta$ and $D=D_\parallel=D_\perp$, when the velocity
dependence is absent we find:
\begin{eqnarray}
\tilde\beta= \beta, \;  \tau=\frac{P\beta}{M^2D}, \;\mu_1(P)=
\frac{1}{M^2D P} \int_0^P d P' \lambda(P')\exp
\left[\frac{\beta}{2M^2D}(P'^2-P^2)\right]. \label{B12}
\end{eqnarray}
Let us introduce the effective temperature $T^\ast=M D/\beta$. If
the stationary distribution is Maxwellian $f_g^0=f_{g0}^T$ with
some temperature $T$ the function $\lambda$ and $\mu_1$ from
Eq.~(\ref{B12}) transforms to
\begin{eqnarray}
\lambda(P)= - \frac{T f_{g0}^{T}}{n}, \,\, \mu_1(P)= -\frac{T
f_{g0}^{T^\ast}}{nM^2D P} \sqrt {\frac{\pi}{2}}
\sqrt{2MT_{eff}}\,\Phi \left[\frac{P}{\sqrt {2M T_{eff}}}\right],
\label{B13}
\end{eqnarray}
where $\Phi(x)$ is the error function and
$1/T_{eff}\equiv(1/T-1/T^\ast)$.

If the Einstein relation $MD=T\beta$ is fulfilled then $T^\ast=T$
and $T_{eff}\rightarrow\infty$. For that case we have
$\mu_1=-Tf_{g0}^{T}/nM^2D$ and the diffusion coefficient in the
coordinate space related with the diffusion in the velocity space
in both 2D and 3D cases by the known equality:
\begin{eqnarray}
D_R=\frac{T^2}{M^2D}.\label{B14}
\end{eqnarray}
For more complicated physical problems, when the Einstein relation
is not applicable, as e.g. for active particles, necessary to use
the general relations derived above. In general necessary also to
use the self-consistent expressions for the friction and diffusion
coefficients in velocity space to find correctly the diffusion in
the coordinate space, because the last one is the functional of
both of them, as it follows from  the general expressions
(\ref{B11}) or (\ref {Ba11}).

However to find $\beta$ and $D$ self-consistently and determine on
that basis the stationary distribution function necessary to find
the probability transition function, which is known on the
microscopical basis only for a restricted type of the momentum
transfer processes \cite{T2002} (Boltzmann-type collisions, some
models of absorption, in particular in dusty plasmas, and some
types of active motion). For many other cases, where the well
developed microscopical model of the physical process is absent,
the approximate non-self-consistent consideration, based on
physical intuition, is only one way to obtain the results.

b) Let us apply now the general relations, found above, to the
model of active particles (e.g. cells) capable to self-motion. The
problem of active friction have been discussed intensively for the
last decade on basis of phenomenological approach (e.g.
\cite{EESS2000}). The microscopical model of active friction for
biological objects was recently suggested in \cite{T2002}.
Generalization of this microscopical consideration for more
complicated and realistic cases, when the "driver" vector of the
object, introduced in this paper, can rotate and orients on some
external gradients of the ambient medium, will be considered
separately. In that case the amplified force, which provides a
negative friction is directed to "driver", although passive
friction is antiparallel to velocity vector, and the active
particle can change the direction of motion.

The PT function for the active particles has been found in
\cite{T2002} by considering physical processes for momentum
transfer from active particles to the ambient medium by internal
energy loss of the inner energy of the grain (cell). Our main
finding is a "singularity" $\beta_\varepsilon(V)\equiv
-K_\varepsilon/V$ in the friction coefficient for active particles
and negative value of $\beta_\varepsilon$. Because there is also
normal positive friction obliged to the surrounding medium
$\beta_0$ the total friction is the sum
$\beta_\Sigma=\beta_0+\beta_\varepsilon(V)$. For the simplest case
$K_\varepsilon$ is V-independent. For more complicated models it
can be a slow function of $V$, as it was shown in \cite{T2002}. We
will consider here only this simplest model. The stationary
distribution function for the active particles is:
\begin{eqnarray}
f_g(V)=  C exp \left \{ - \frac{\beta_0}{2D_\Sigma}\left(V -
\frac{K_\varepsilon}{\beta_0}\right)^2 \right\},\label{B15}
\end{eqnarray}
where $D_\Sigma=D_0+D_\varepsilon$ is the total diffusion, which
is a constant if $K_\varepsilon$ is V-independent. The
coefficients in Eq.~(\ref{B15}) are found on the basis of a
kinetic consideration \cite{T2002}. The characteristic parameters
determined above in the case under consideration can be taken
$\beta_c\equiv\beta_0$, $V_0=K_\varepsilon/\beta_0$, $D_c\equiv
D_\Sigma$ and $d(V)\equiv1$. The physical reason for the maximum
of the distribution function at $V=K_\varepsilon/\beta_0$ is
evident. It is due to the balance of the normal friction and
self-accelerating forces. For self-motion of grains (cells) for
example in a gas of hard spheres we can use for $\beta_0$ and
$D_0$ the expressions (\ref{beb2}, \ref{be4}). A similar
$V$-dependence of the distribution function was found
experimentally and phenomenologically for cells \cite{SG93,SFG94}.
Using Eq.(\ref{Ba11}, \ref{Bc11}) we find (for $s=3$):
\begin{eqnarray}
D_R = \frac {K_\varepsilon^4}{3 \beta_0^4 D_\Sigma}\,\digamma
\left( \frac{K_\varepsilon^2}{\beta_0 D_\Sigma}\right)\label{Bc11}
\end{eqnarray}
In many other cases, when the microscopic theory of active motion
is not developed yet the velocity independent diffusion also seems
the suitable approximation \cite{EESS2000}. For the all these
models Eq.~(\ref{B11}) can be simplified:
\begin{eqnarray}\
D_R = -\frac {1}{s M^3D} \int d^s P \frac {P}{f_g^0(P)}\int_0^P d
P' \lambda(P')f_g^0(P'), \label{B16}
\end{eqnarray}
If the characteristic constant value for the function $\beta(P)$
is $\beta_c$ and the characteristic momentum is $P_0$ (e.g. for
the active particles the value of momentum, where $\beta(P_0)=0$)
we can introduce the undimensional function $\phi(P/P_0)$ by the
equality $\beta(P)=\beta_c \phi (P/P_0)$ and rewrite
Eq.~(\ref{B11}) in the form:
\begin{eqnarray}
D_R = -\frac{P_0^4}{s M^4D} \Omega(\Lambda),\label{B17}
\end{eqnarray}
where the parameter $\Lambda=P_0^2\beta_c/MD$ and $\Omega\equiv
\,Y/N$ is the the ratio of the integrals:
\begin{eqnarray}
Y= \int d^s \tau\;\tau \int_0^{\tau} d \eta\,\int_\eta^\infty d
\xi \xi f_g^0(\xi)\exp \left[\Lambda \int_\tau^\xi d
\omega\phi(\omega)\right], \label{B18}\\
N=\int d^s \xi f_g^0(\xi).\label{B19}
\end{eqnarray}
For the self-consistent scheme the distribution function itself,
according to Eq.~(\ref{fgr}), is equal to:
\begin{eqnarray}
f_g^0(\xi)=C \exp \left[-\Lambda \int_0^\xi d\rho
\phi(\rho)\right].\label{B20}
\end{eqnarray}

\section{Diffusion in the coordinate space on the basis of
master-type equation}
Consideration of diffusion in the coordinate space and relation
between the diffusion in coordinate and velocity space have been
done in Section 3 on the basis of the Fokker-Planck kinetic
equation. At the same time this approach is not enough for
consideration of some special types of space diffusion, when the
PT function has a specific form, in particular, posses the slow
decreasing tails in the coordinate space, when the
Fourier-components for such PT functions are absent. There are
many very important processes, where different types of the
anomalous diffusion exist \cite{MK2000}. This anomalous diffusion
can be related not only with the coordinate, but also, in general,
with time-dependent PT function. Usually the problem of anomalous
diffusion is considering on the basis of fractional
differentiation. This basis provides the universal description for
probability distribution of grain at large distances. At the same
time to describe all the distances we suggest to use the approach
of PT function, when the fractional differentiation is not present
at all. Naturally the results for the power-type PT functions are
the same. But in general case, the PT function approach gives
general and simple description for all interesting cases with a
wide class of short-range behavior and a long distance tails in
the coordinate space. We also formulate the time-dependent
equation for anomalous diffusion, which gives useful general
description for many applications. These applications will be
considered in details in the separate paper.

Let us consider the diffusion in the coordinate space on the basis
of master equation, which describes the balance of the grains
incoming and outcoming the point $r$ in the moment $t$. The
structure of this equation is formally similar to the master
equation in the momentum space Eq.~(\ref{CK}). Naturally, for the
coordinate space there is no the conservation law, similar to the
momentum space:
\begin{equation}
\frac{df_g({\bf r},t)}{dt} = \int d{\bf r'} \left\{W ({\bf r, r'})
f_g({\bf r', t}) - W ({\bf r', r}) f_g({\bf r},t) \right\}.
\label{DC1}
\end{equation}
The probability transition $W({\bf r, r'})$ describes probability
for a grain to transfer from the point ${\bf r'}$ to the point
${\bf r}$ per unit time. We can rewrite this equation in the
coordinates ${\bf \rho= r'- r}$ and ${\bf r}$ as:
\begin{equation}
\frac{df_g({\bf r},t)}{dt} = \int d{\bf \rho} \left\{W ({\bf \rho,
r+\rho}) f_g({\bf r+\rho, t}) - W ({\bf \rho, r}) f_g({\bf r},t)
\right\}. \label{DC2}
\end{equation}
If we suggest that the characteristic displacements are small and
expand Eq.~(\ref{DC2})  we arrive to the "Fokker-Planck" form of
the equation for the density distribution $f_g({\bf r},t)$
\begin{equation}
\frac{df_g({\bf r},t)}{dt} = {\partial \over {\partial r_\alpha}}
\left[ A_\alpha ({\bf r}) f_g({\bf r},t) + {\partial \over
{\partial r_\beta}} \left( B_{\alpha\beta}({\bf r}) f_g({\bf r},t)
\right)\right], \label{DC3}
\end{equation}
The coefficients $A_\alpha$ and $B_{\alpha \beta}$ describe the
acting force and diffusion, respectively, can be written as the
functionals of the probability function (PT) in the coordinate
space $W$ in the form:
\begin{equation}
A_\alpha({\bf r}) = \int d^s \rho \rho_\alpha W({\bf \rho, r})
\label{DC4}
\end{equation}
and
\begin{equation}
B_{\alpha\beta}({\bf r})= \frac{1}{2}\int d^s\rho \rho_\alpha
\rho_\beta W({\bf\rho, r}). \label{DC5}
\end{equation}
For the isotropic case the the probability function depends from
${\bf r}$ the modulus of $\rho$. For homogeneous medium, when
$r$-dependence of the PT is absent the coefficients $A_\alpha=0$
and the diffusion coefficient is constant
$B_{\alpha\beta}=\delta_{\alpha\beta}B$, where B is the integral
\begin{equation}
B = \frac{1}{2s}\int d^s \rho \rho^2 W(\rho). \label{DC6}
\end{equation}

This consideration cannot be applied to the specific situations,
when the integral in Eq.~(\ref{DC6}) is infinite. In that case we
have investigate the general transport equation (\ref{DC1}). We
will consider the problem for homogeneous and isotropic case, when
PT function depends only from modulus $|\rho|$. By
Fourier-transformation we arrive to the form of Eq.~(\ref{DC1}):
\begin{equation}
\frac{df_g({\bf k},t)}{dt} = \int d^s \rho  \left[\exp (i {\bf
k}\rho) -1 \right]W(|\rho|) f_g({\bf k},t)\equiv X({\bf
k})f_g({\bf k},t), \label{DC7}
\end{equation}
where $X({\bf k})\equiv X(k)$. Let us consider the simple form of
PT function with a power dependence from distance $W(\rho)=
C/|\rho|^\alpha$, where $C$ is a constant. For one-dimensional
case we find:
\begin{equation}
X(k)\equiv -4 \int_0^\infty d\rho\, sin^2
\left(\frac{k\rho}{2}\right)W(\rho )= - 2^{3-\alpha}C
|k|^{\alpha-1} \int_0^\infty d\zeta
\frac{sin^2\zeta}{\zeta^\alpha}. \label{DC8}
\end{equation}
For the values $1<\alpha<3$ this function is finite and equal
\begin{equation}
X(k) = - 2C \Gamma(1-\alpha) |k|^{\alpha-1}\,\cos
\frac{(1-\alpha)\pi}{2} , \label{DC9}
\end{equation}
where $\Gamma$ is the Gamma-function. At the same time the
integral (\ref{DC6}) for such type of PT functions is infinite,
because usual diffusion is absent. The considered procedure for
the simplest cases of power dependence of PT function is
equivalent to the equation with fractional space differentiation
\cite{S1987,F1994,BS2002}:
\begin{equation}
\frac{df_g(x,t)}{dt} = C\Delta^{\mu/2}f_g(x,t),\label{DC10}
\end{equation}
where $\Delta^{\mu/2}$ is a fractional Laplacian, a linear
operator, whose action on the function $f(x)$ in Fourier space is
described by $\Delta^{\mu/2}f(x)=-(k^2)^{\mu/2}f(k)=-|k|^\mu
f(k)$. In the considered above case $\mu\equiv(\alpha-1)$, where
$0<\mu<2$. At the same time for more general PT functions, which
are (for arbitrary values $\rho$) are not proportional to the
-$\alpha$th power of $\rho$, the method described above is also
applicable, although the fractional derivative is not exists.

For the cases of purely power dependence of PT the non-stationary
solution for density distribution describes so-called
super-diffusion (or Levy flights). The solution of
Eq.~(\ref{DC10}) in Fourier space reads:
\begin{equation}
f_g(k,t) = \exp (-C|k|^\mu t),\label{DC11}
\end{equation}
which in coordinate space corresponds to a so-called symmetric
Levy stable distribution:
\begin{equation}
f_g(x,t) = \frac {1}{(kt)^{1/\mu}} L\left[\frac{x}{(kt)^{1/\mu}};
\mu, 0 \right]. \label{DC12}
\end{equation}
For general case as it follows from Eq.~(\ref{DC7})
\begin{equation}
f_g(k,t) = C_1 \exp [X(k)t],\label{DC12}
\end{equation}
with some constant $C_1$.

Let us turn to the three-dimensional case for the PT function
equal $W=C/|\rho|^\alpha$ with a constant $C>0$. Then the function
$X(k)$ is equal
\begin{equation}
X(k) \equiv 4\pi \int_0^\infty d\rho \rho^2 \left(\frac{\sin
(k\rho)}{k\rho}-1\right)W(\rho)= 4\pi  C k^{\alpha-3}\int_0^\infty
d\zeta\frac{1}{\zeta^{\alpha-2}}\left(\frac{\sin\zeta}{\zeta}-1\right)
\label{DC13}
\end{equation}
This integral is finite for the values $3<\alpha<5$. If we return
to the variable $\mu\equiv\alpha-s$ (s=3 for 3-dimensional case
and equal to the power of $k$ in the function $X(k)$) we find the
same limitation as in one-dimensional case $0<\mu<2$, which
characterizes the fractional derivative.

Naturally, consideration on the basis of PT function given above,
permits to avoid the fractional differentiation method and to
consider more general physical situations of the non-power
probability transitions. Let us consider that for a simple
example. Taking (for one-dimensional case) the PT function
$W(\rho)$ in the form
\begin{equation}
W(\rho) = C \frac{1-\exp [-\sigma\rho^p]}{\rho^\alpha},
\label{DC14}
\end{equation}
with $p>0$, we arrive (e.g. for one-dimensional case) to the
function $X(k)$:
\begin{equation}
X(k)= - 2^{3-\alpha}C |k|^{\alpha-1} \int_0^\infty d\zeta
\frac{\left\{1-\exp [-\sigma (2\zeta/|k|)^p\;]\right\}
sin^2\zeta}{\zeta^\alpha}\equiv -2^{3-\alpha}C
|k|^{\alpha-1}Y(\sigma/|k|^p,\alpha). \label{DC15}
\end{equation}
As easy to see the function $Y(\sigma/|k|^p,\alpha)$ is finite for
$1<\alpha<p+3$, because for the small values of distances for
$p>0$ divergence is suppressed also for some powers $\alpha>3$.
Simple calculation for $\alpha=2$ and $p=1$ leads to the result,
which cannot be found by the fractional differentiation method
\begin{equation}
Y(\sigma/|k|,2)= \frac{\pi}{2}-\arctan
(|k|/\sigma)+\frac{\sigma}{2|k|} \ln \left[1+k^2/\sigma^2\right].
\label{DC16}
\end{equation}
The asymptotic behavior of the function $X(k)$ for
$k\rightarrow\infty$ is similar, as it follows from
Eq.~(\ref{DC16}), to the case $W(x)=c/\rho^\alpha$. Therefore the
universal behavior of $X(k)$ is provided by asymptotical
properties of PT function for $1<\alpha<3$.

For the normal Gaussian diffusion (see e.g. Eqs.~(\ref{DC3}),
(\ref{DC6})) for arbitrary $s$ the function is equal $X(k)=-Bk^2$.
It leads to the Gaussian distribution in the coordinate space and
to the usual time dependence of the square-mean displacement
$\langle r^2\rangle\sim Bt$. For different types of anomalous
diffusion this relation is violated and the dependence $\langle
r^2(t)\rangle$ has to be calculated on the basis of concrete
anomalous distribution, which is determined by the respective
equation for $f_g({\bf r},t)$. As we already demonstrated and will
show below the crucial function, which determines the process is
the PT function $W$.

We also would like to mention in the connection of the problem of
generalized description of diffusion that the time dependence of
the density distribution for the time dependent PT function can be
very different from the classical diffusion. This kind of problems
relates to the class of the stochastic transport, which is very
popular in modern physics and describes so called subdiffusive
behavior \cite{SM1975}, e.g. photoconductivity in strongly
disordered and glassy semiconductors or the resonance radiative
transfer in a plasma \cite{ZC2002}, which posses many other
applications. Here we only shortly formulate the problem, leaving
it for the detail consideration in the next paper. For this
purpose we formulate a more general transport equation for density
distribution:
\begin{equation}
f_g({\bf r},t) = f_g({\bf r},t=0)+ \int_0^t d\tau \int d{\bf r'}
\left\{W ({\bf r, r'},\tau, t-\tau) f_g({\bf r',\tau}) - W ({\bf
r', r},\tau, t-\tau) f_g({\bf r},\tau) \right\}. \label{DC17}
\end{equation}
For the case of stationary PT function we return to
Eq.~(\ref{DC1}). If there is no memory, but PT function is a
function of current time $\tau$ we arrive to the equation more
general than Eq.~(\ref{DC1}), which describes the density
evolution, with prescribed time dependence of PT. In particular in
that case for slow space dependence of the function $W$ we find
the equation for diffusion similar to Eq.~(\ref{DC3})-(\ref{DC5})
with time-dependent coefficients $A_\alpha({\bf r},t)$ and
$B_{\alpha\beta}({\bf r},t)$, which are calculated on the basis of
the PT function $W({\bf \rho, r},t)$.

If the system posses memory, then, in the simplest case, when the
function $W$ can be expanded in spirit of Fokker-Planck
approximation in the coordinate space we arrive to the following
form of Eq.~(\ref{DC17})
\begin{equation}
f_g({\bf r},t) = f_g({\bf r},t=0)+ \int_0^t d\tau {\partial \over
{\partial r_\alpha}} \left[ A_\alpha ({\bf r},t-\tau) f_g({\bf
r},\tau) + {\partial \over {\partial r_\beta}} \left(
B_{\alpha\beta}({\bf r},t-\tau) f_g({\bf r},\tau) \right)\right].
\label{DC18}
\end{equation}

If the function $W$ posses memory and depends only from the
difference $t-\tau$, but cannot be expanded in the coordinate
space as in the case (\ref{DC18}) we can use the
Laplace-transformation in time to find:
\begin{equation}
f_g({\bf r},z) = \frac{f_g({\bf r},t=0)}{z}+ \int d{\bf r'}
\left\{W ({\bf r, r'},z) f_g({\bf r'},z) - W ({\bf r', r},z)
f_g({\bf r},z) \right\}. \label{DC19}
\end{equation}
To arrive to this equation we suggested that due to fast
convergence the upper limit $t$ in the integral $\int_0^t d\tau
W(\tau)\exp(-z\tau)$ with Im $z>0$ can be changed on infinity and
determine $W ({\bf r', r},z)$ by the equality:
\begin{equation}
W ({\bf r', r},z)=\int_0^\infty d\tau W({\bf r',
r},\tau)\exp(-z\tau)\label{DC20}
\end{equation}
(The same approach can be used to simplify the Fokker-Planck-type
equation (\ref{DC18})). For the space homogeneous case
Eq.~(\ref{DC19}) can be Forier-transformed and rewritten in the
$k,z$ variables:
\begin{equation}
f_g({\bf k},z) = \frac{f_g({\bf k},t=0)}{z\left[1-X({\bf
k},z)\right]}, \label{DC21}
\end{equation}
where
\begin{equation}
X({\bf k},z)= \int d^s \rho  \left[\exp (i {\bf k}\rho) -1
\right]W(|\rho|,z).\label{DC22}
\end{equation}
If the PT function is time-independent $W(|\rho|,z)=W(|\rho|)/z$
and $f_g({\bf k},t=0)=$const. we return, as easy to see, to the
case of abnormal diffusion considered above in
Eqs.~(\ref{DC7})-(\ref{DC16}). For a generale multiplicative form
of PT function $W(|\rho|,\tau)=W_1(|\rho|)W_2(\tau)$ the function
$X(k,z)$ is the product:
\begin{equation}
X({\bf k},z)\equiv X_1(k)X_2(z)= \int d^s \rho  \left[\exp (i {\bf
k}\rho) -1 \right]W_1(|\rho|)\cdot\int_0^\infty d\tau
W_2(\tau)\exp(-z\tau).\label{DC23}
\end{equation}
For the power $\tau$-dependence $W_2=C/\tau^\gamma$ with
$\gamma<1$ the integral $X_2(z)$ is equal
\begin{equation}
X_2(z)= \frac{1}{z^{1-\gamma}}\,\Gamma(1-\gamma).\label{DC24}
\end{equation}
The distribution function (\ref{DC21}) in $k,z$ space in that case
reads
\begin{equation}
f_g({\bf k},z) = \frac {f_g({\bf k},t=0)}{\left[z-z^\gamma\,
\Gamma(1-\gamma) X_1(k)\right]}, \label{DC25}
\end{equation}

The detail consequences and physical applications of more
complicated dependence $W(|\rho|,z)$
from $z$ will be considered in the separate paper.

\section{Generalized friction}
Let us consider now the generalized expression for the friction
force ${\bf F}_f$ on the basis of the Fokker-Planck equation for
charged grains. As it was shown in \cite {T2002} for scattering of
the ion stream with velocity ${\bf u}$ by moving grain with
momentum ${\bf P_0}$ the of PT function $\tilde w_s({\bf P_{0},y,
q})$ satisfies the equality:
\begin{equation}
\tilde w_s({\bf P_{0},y, q})= w_s({\bf P_{0}-y, q}), \label{ws}
\end{equation}
where $w_s({\bf P, q})$ is the PT for scattering of an isotropic
system of ions by moving grains and ${\bf y}=M{\bf u}$. The
generalized friction force for one grain can be represented
through the friction coefficient:

\begin{equation}
{\bf F}_{fs}({\bf P_0- y}) = - ({\bf P_0- y})\, \beta_s
(\left|{\bf P_0- y}\right|). \label{Ff4}
\end{equation}
In particular, Eq.~(\ref{Ff4}) describes both limiting cases: the
friction force itself (for ${\bf u}=0$) and the drag force itself
(for ${\bf P}_{0}=0$), acting on the grain due to ion scattering
with a non-zero driven velocity. For two limiting cases of the
friction force itself ${\bf F}_{f0s}$ for the grain with momentum
${\bf G}$ and immobile ions and for the opposite case - ion drag
itself ${\bf D}_{\emph{i\,s}}$ with the ion velocity ${\bf u}=
{\bf G} /M$ there is natural relation ${\bf F}_{f0s}({\bf G})=
-{\bf D}_{\emph{i\,s}}({\bf G})$. This picture leads to the (for
practical applications important) generalization for a few species
of the lighter particles, which we consider below. The particular
result for  ${\bf D}_{\emph{i\,s}}$ for the case $\mid{\bf u}\mid
\ll v_{T_i}$ and the Coulomb scattering cross-section can be
written in the form:
\begin{eqnarray}
{\bf D}_{\emph{i\,s}}= 2A_0 M {\bf u}\Gamma^2 \ln\Lambda,
\label{ad11}
\end{eqnarray}
where $A_0=\frac{\sqrt{2\pi}}{3} (m_i/M) a^2 n_i v_{T_i}$,
$v_{T_i}\equiv\sqrt{T_i/m_i}$, and $a$ is the radius of grain. For
dusty plasmas usually the parameter $\Gamma \equiv e^2 Z_g Z_i /a
T_i\gg1$, and the Landau logarithm $\ln\Lambda$ is given in \cite
{KI2002}, \cite {STZ2003}.

For the collecting PT function $\tilde w_c({\bf P,y,q})$, based on
the existing collecting collision integral \cite {IG1998}, a
representation similar to Eq.~(\ref{ws}) is not valid
\cite{T2002}. Instead of Eq.~(\ref{ws}) we have:
\begin{equation}
\tilde w_c({\bf P,y,q}) = w_c ({\bf P-y,q + \emph{m}
u}),\label{ad13}
\end{equation}
where $w_c ({\bf P, q})$ is the PT for the collecting process (in
the case under consideration for the ions) in the isotropic ion
system $({\bf u}=0)$, e.g., the Maxwellian one. The collecting
part of the ion friction force for $\eta\equiv v^2/ 2v^2_{\rm
Ti}\ll1$ and arbitrary $\Gamma$ has a form \cite {TZ2002}, \cite
{T2002}:
\begin{eqnarray}
{\bf F}_{fc}= - M {\bf V} \beta(\eta,\Gamma) = -2M {\bf V} A_0
\left[ 1-\Gamma - {\eta\over5} (1-3\Gamma)\right]. \label{Na1}
\end{eqnarray}
For $\Gamma>1$ this formula describes a negative collecting
friction ${\bf F}_{fc}$ parallel to ${\bf V}$. At the same time
the collecting ion drag force ${\bf D}_{\emph{i\,c}}$ (derived in
\cite {KI2002} and rigorously justified later in \cite {T2002})
for $\xi\equiv u^2/ 2v^2_{\rm Ti}\ll1$ and arbitrary $\Gamma$ can
be written as:
\begin{eqnarray}
{\bf D}_{\emph{I\,c}} = 8A_0 M {\bf u} \left[ 1+ \frac
{\Gamma}{2}\right]. \label{Na2}
\end{eqnarray}
The difference of the modulus of forces in Eqs.~(\ref{Na1}) and
(\ref{Na2}) (for the case $u=V$) is the reflection of the
collecting PT property, expressed by Eq.~(\ref{ad13}) and the
structure of the collision collecting integral in dusty plasmas,
when all the ions which cross the grain surface are absorbed and
therefore completely transfer their momentum to the grain,
although the mass transfer is considered as negligible. The
problem of the mass transfer has been considered in \cite {T2001}-
\cite {ITME2002} where a mass kinetic variable for dusty plasmas
was introduced. The explicit solutions obtained in these papers
were non-stationary. For RF dusty plasmas we are interested to
find stationary solutions for the grain distribution function.
These solutions take into account the surface recombination of the
ions that bombarded the grain and remove the created atoms to
ambient plasmas. Such a kinetic model (very close to many real
experiments in RF discharge with dust) is in the process of
being developed.

\section{Negative generalized friction for the case of ion
flow}
Let us consider now the important generalization of the above
approach for the case of two species of light particles, which
transfer momentum to the grains. We assume an ion stream with
velocity ${\bf u}$, which can be scattered and absorbed by the
grains (with velocity ${\bf V}$) and an atomic subsystem with an
isotropic velocity distribution function, e.g., a Maxwellian
distribution. Then by using the method developed in \cite {T2002}
and assuming that for the considering model the equality
(\ref{ws}) is satisfied, we arrive at the total force acting on
the grains as expressed through the total friction coefficient
$\beta_t(V)$:
\begin{equation}
{\bf A}_{\alpha} (P,y) = {\bf P}_{\alpha}\beta_t(V), \label{Ff5}
\end{equation}
\begin{equation}
\beta_t(V)= \beta_a(V) - \frac{y}{P}\beta_i(y). \label {Ff6}
\end{equation}
Using the approximations for atomic and ion friction coefficients
$\beta_a(V)\simeq\beta_a(0)$, where $\beta_a(0)= 8
\frac{\sqrt{2\pi}}{3} (m_a/M) a^2 n_a v_{T_a}$ (scattering of a
point atom on the hard sphere grain) and $\beta_i(V)\simeq 2A_0
\Gamma^2\ln\Lambda$. This approximation for $\beta_i(V)$ is valid
for dominant ion scattering drag \cite{TEIT2003} and permits to
avoid the problem of possible negative friction related with the
mechanism of ion absorption, discussed in \cite {TZ2002}. Such a
way we arrive, for the simplest case, when the vectors of ${\bf
P}$ and ${\bf u}$ are parallel, at the picture typical for
negative effective total friction. The total friction coefficient
\begin{equation}
\beta_t(V)= \beta_a(0) \left[1 - \frac{u m_i n_i v_{T_i}}{2V m_a
n_a v_{T_a}}\Gamma^2\ln\Lambda \right] \label {Ff67}
\end{equation}
becomes negative for small grain velocities.

\section{Conclusions}
The paper devoted to consideration velocity-dependent friction and
diffusion coefficients in the velocity space and diffusion,
including anomalous one, in the coordinate space. We established
the general relations, which determine these coefficients
(including tensorial structure of diffusion) and showed, that this
relations exist not only for the Fokker-Planck equation, which
applicability is restricted, at least for the Boltzmann-type of
collisions, by low range of grain velocity. On the basis of master
equation with the prescribed PT function we showed validity of
these general relations for arbitrary grain velocities.

We also considered the application of the PT function method to
the ion absorption collisions in dusty plasmas, where due to the
specific structure of the absorption cross-section and PT function
the negative friction appears for a domain of low grain velocity.
Necessary to mention that this result is valid for the considered
model \cite{TZ2002}, but for application of the results to the
real dusty plasmas, where the additional processes of surface ion
recombination and atom remove to the plasma are exist, the
elaborated models have to be developed.

The important question about relation between the diffusion
coefficients in the velocity and coordinate spaces is solved for
the Fokker-Planck equation. For open systems, or for the systems
with velocity dependent coefficients the relation with coordinate
diffusion is essentially more complicated, than in the usual case
of time independent coefficients, connected by the Einstein
relation.

The problems of anomalous diffusion are considered on the basis of
PT for the master-type equation in the coordinate space.
Consideration presented above permits to avoid the method of
fractional space differentiation (coinciding with the results
obtained by this method in the particular cases of power-type
dependencies of PT) and to extend the results for a wide class of
PT functions. The extension of this approach is used to formulate
the general integral equation for distribution function in the
coordinate space, applicable to description of very different
types of time-dependent normal and anomalous diffusion. This
results, as we trust, can be important for very different
applications.

In many cases the PT function in the velocity space depends from
some other vectors, except grain velocity. It can be ion driven
velocity, for example, in dusty plasmas, or "driver" vector of
cells in biological systems \cite{T2002}, or some additional
vector for colliding inelastic grains etc. For these cases we have
consider the methods of PT function determination and calculation
the generalized friction and diffusion. Some development of these
approach is also presented in the paper in application to
scattering and absorption of ions by grains in dusty plasmas.

For the case of scattering of ions, which have some driven
velocity, by grains, moving in the slightly ionized dusty plasmas
with essential friction between grains and atomic component we
show manifestation of negative friction for low grain velocities.
This process can be easily realized for some parameters of dusty
plasma, providing acceleration and heating of grains.

\acknowledgments
Authors thanks W. Ebeling, A.M.Ignatov and I.M. Sokolov for
valuable discussions of some problems, reflected in this work.

\end{document}